\documentclass[prl,aps, twocolumn, floatfix, superscriptaddress]{revtex4}

\usepackage{graphicx, epsfig}
\usepackage{amssymb,amsmath}

\def \hf{\tfrac{1}{2}}    

\def \ord{\mathcal{O}}
\def \ra{\rightarrow}   

\def\lba{\left(}    \def\rba{\right)}

\newcommand{\ket}[1]{\left|{#1}\right.\rangle}

\DeclareMathOperator{\tr}{tr}

\begin{document}

\title{Entanglement entropy in fermionic Laughlin states}

\author{Masudul Haque} 
\affiliation{Institute for Theoretical Physics, Utrecht University,
the Netherlands}

\author{Oleksandr Zozulya} 
\affiliation{Institute for Theoretical Physics, University of
Amsterdam, 
the Netherlands}

\author{Kareljan Schoutens} 
\affiliation{Institute for Theoretical Physics, University of
Amsterdam, 
the Netherlands}

\date{\today}

%
%

\begin{abstract}

We present analytic and numerical calculations on the bipartite
entanglement entropy in fractional quantum Hall states of the
fermionic Laughlin sequence.  The partitioning of the system is done
both by dividing Landau level orbitals and by grouping the fermions
themselves.  For the case of orbital partitioning, our results can be
related to spatial partitioning, enabling us to extract a topological
quantity (the `total quantum dimension') characterizing the Laughlin
states.  For particle partitioning we prove a very close upper bound
for the entanglement entropy of a subset of the particles with the
rest, and provide an interpretation in terms of exclusion statistics.


\end{abstract}




\maketitle

\emph{Introduction} ---
While the renewed attention in quantum entanglement has mostly been in
the context of quantum computing, there is now also growing interest
in using entanglement measures for characterizing quantum
many-particle states.  The measure we focus on is the entanglement
entropy, which is defined by partitioning the system under question
into two blocks $A$ and $B$, and using the reduced density matrix of
one part (e.g., $\rho_A = \tr_B\rho$ obtained by tracing over $B$
degrees of freedom) to calculate the von Neumann entropy $S_A =
-\tr[\rho_A\ln\rho_A]$.  Generally speaking, the entanglement entropy
scales with the size of the boundary between the partitions
\cite{Srednicki_PRL93}.  Prefactors, logarithmic corrections and
subleading terms in this basic relationship can all contain important
physical information concerning the many-body state.

By now the entanglement entropy has been studied extensively for
one-dimensional spin systems, where the variation of $S_A$ as a
function of the size of block $A$ is a sensitive probe into the nature
of the ground state \cite{Entng-in-spin-chains}.  Entanglement
measures have also been shown to contain revealing behavior in the
vicinity of quantum critical points \cite{Entng-n-QCP} and shown to
have fundamental implications for the efficiency of numerical
simulation methods for quantum many-body states
\cite{Entng-n-computation}.

Studies of entanglement in higher-dimensional and itinerant systems
has been less thorough but have now also started to attract significant
attention (e.g., \cite{Entng-high-D, Hamma-Ionicioiu-Zanardi_PRA05,
Preskill-Kitaev_PRL06, Levin-Wen_PRL06}).  The
case of two dimensions is particularly intriguing because of the known
existence (and further unconfirmed possibilities) of topologically
ordered phases.  It is natural to speculate that entanglement might
give a handle on topological properties of quantum states, because the
basic intuition about topologically ordered phases is the presence of
intricate correlations not easily captured by local observables or
traditional correlation functions.
Indeed, Refs.~\cite{Preskill-Kitaev_PRL06} and \cite{Levin-Wen_PRL06}
have recently presented the following theorem concerning topologically
ordered states: if $L$ is the boundary between the two blocks, the
entanglement entropy scales as $S_A = \alpha{L} -\gamma
+\ord(L^{-1})$.
As usual the scaling law applies to situations where $A$ is large and
the total system is infinite.
The quantity $\gamma$ (the topological entanglement entropy) is the
logarithm of a quantity known as the \emph{total quantum dimension}.
For a state with anyonic excitations, the quantum dimensions
characterize the growth rate of the Hilbert space with anyon
number. For fermionic fractional quantum Hall states in the Laughlin
sequence with filling fraction $\nu=1/m$, the topological entanglement
entropy is $\gamma = \hf\ln{m}$.

In this Letter, we present a detailed study of the entropy of
entanglement in the fermionic Laughlin states.  These are the
best-known many-particle states with topological order.  We carefully
choose useful methods of dividing quantum Hall states into $A$ and $B$
blocks.  We provide results on the entropy defined by partitioning the
Landau-level orbitals in the system into blocks, as well as that
obtained by dividing the particles themselves into $A$ and $B$
subsets.  The first method allows us, upon extrapolation to the
thermodynamic limit, to extract the $\gamma$ parameter of
Refs.~\cite{Preskill-Kitaev_PRL06, Levin-Wen_PRL06}.  To our best
knowledge, this is the first example where the results of
Refs.~\cite{Preskill-Kitaev_PRL06, Levin-Wen_PRL06} are used to
compute the total quantum dimension of an experimentally realized
topologically ordered medium, directly from a microscopic description.

Very restricted cases of entanglement in Laughlin states, involving
one-orbital or one-particle blocks, have appeared previously in the
literature, in Refs.~\cite{Shi_JPhys04, Tsinghua_FQHE_PRA02}.  The
entanglement entropy in some other topologically ordered states has
been studied in Ref.~\cite{Hamma-Ionicioiu-Zanardi_PRA05}.

\emph{Orbital versus particle entanglement} ---
We first explain our choice of methods used to partition the Laughlin
system into blocks $A$ and $B$.  The issue is non-trivial in the case
of itinerant systems \cite{Shi_JPhys04, Dowling-etal_PRA06}, where,
unlike spin systems, the particles are not each fixed in its own site.
Although this point has not been stressed in the literature, conformal
field theory results on entanglement scaling \cite{Cardy_JStatMech04}
actually pertain to the blocking of \emph{space} rather than the
particles or spins themselves.  For quantum Hall systems, we find a
description of the system in terms of the magnetic orbitals more
natural than a spatial description.  Thus we will partition the
orbitals into two sets $A$ and $B$ and calculate the entanglement
between them.
In fact, there is a close relationship between orbital and spatial
partitioning, since the choice of the first $l_A$ orbitals as the $A$
block corresponds approximately to choosing a disk-shaped $A$ block
with radius proportional to $\sqrt{l_A}$ in real space.  
The scaling law of Refs.~\cite{Preskill-Kitaev_PRL06, Levin-Wen_PRL06}
thus translates to $S_{l_A} = c_1\sqrt{l_A} -\gamma + \ord(1/l_A)$.

In addition, we also consider the perhaps more obvious method of
partitioning the particles themselves.  In this case, the subsets $A$
and $B$ no longer correspond to connected regions in space.  We find
that the analysis of particle entanglement also reveals subtle
correlation effects.

\emph{Quantum Hall states on a sphere} ---
To describe the fractional quantum Hall states, we use the
representation of Refs.~\cite{Haldane_FQHE_PRL83,
ArovasAuerbachHaldane_PRL88} in which the fermions are placed on a
sphere containing a magnetic monopole.  The magnetic orbitals in the
lowest Landau level are then represented as angular momentum orbitals;
for $N$ particles in the Laughlin state $\nu=1/m$, the total angular
momentum is half the number of flux quanta, $L={\hf}N_{\phi}$ with
$N_{\phi}=m(N-1)$.  The $N_{\phi}+1$ orbitals are labeled either $l=0$
to $N_{\phi}$ or $L_z=-L$ to $+L$.  The ``filling'' acquires the usual
meaning $\nu = N/N_{\phi}$ only in the thermodynamic limit.
The orbitals are each localized around a ``circle of latitude'' on the
sphere, with the $l=0$ orbital localized near one ``pole.''  
For orbital partitioning, we define block $A$ to be the first $l_A$
orbitals, extending spatially from one pole out to some latitude.
In the thermodynamic limit, this is equivalent to
having a disk-shaped block $A$; since each orbital $l$ is associated
with a wavefunction of the form $z^le^{-|z|^2}$ in usual complex
coordinate language, a disk with $l_A$ orbitals has radius
$\sim\sqrt{l_A}$.

In this representation the Laughlin wavefunctions are expressed in
terms of Schwinger boson operators $a_i^{\dagger}$, $b_i^{\dagger}$ as
$\prod_{i<j} (a_i^{\dagger}b_j^{\dagger}-b_i^{\dagger}a_j^{\dagger})^m
\ket{0}$.  Each term in the expansion can be interpreted in terms of
orbital occupations.  For example, in the expansion for the $N=3$,
$m=3$ wavefunction, the term involving $(a_1^{\dagger})^1
(a_2^{\dagger})^2 (a_3^{\dagger})^6$ corresponds to having the three
particles in orbitals 1, 2, and 6 respectively.  Together with its
permutations, this state reads in ``orbital'' notation
[0,1,1,0,0,0,1].

\emph{Analytic results for orbital partitioning} ---
We first consider the simplest Laughlin state, $m=3$.  With only the
$l=0$ orbital included in partition $A$ ($l_A=1$), the reduced density
matrix $\rho_A$ is $2\times{2}$ and diagonal, and has eigenvalues
$N/(3N-2)$ and $(2N-2)/(3N-2)$.  The entanglement entropy converges in
the large-$N$ limit as $S_{l_A=1} = \ln(3/2^{2/3}) + \ln{2^{2/9}}/N +
\ord(N^{-2})$.

The case of the first two orbitals in the $A$ block ($l_A=2$) is also
exactly solvable.  The basis for $\rho_A$ is the set [0,0], [1,0],
[0,1], [1,1]; however the last one can be dropped because the two
lowest orbitals are never simultaneously occupied in the $m=3$
wavefunctions \cite{Haldane_FQHE_PRL83}.  The reduced density matrix
is still diagonal, with a pair of eigenvalues $N/(3N-2)$ and a single
$(N-2)/(3N-2)$.  The thermodynamic limit is $S_{l_{A}=2} = \ln{3} +
4/(9N^2) + \ord(N^{-3})$.

The case of $l_A=3$ is already more difficult.  The reduced density
matrix, now of size $5\times5$, is still diagonal; the elements are of
the form (up to normalization) $N$, $\alpha_N$, $\alpha_N$,
$(N-\alpha_N)$, and $(N-2-\alpha_N)$.  We find $\alpha_3= 1$,
$\alpha_4= 92/51$ and $\alpha_5= 505/203$, etc., but a general form is
not obvious; thus we lack an analytic expression for $S_{l_{A}=3}$ in
the thermodynamic limit.

We now turn to a general Laughlin state of filling $\nu=1/m$.  For
this state there is at most one fermion in the first $\hf(m+1)$
orbitals \cite{Haldane_FQHE_PRL83}; this makes $S_{l_A}$ analytically
tractable for $l_A \leq \hf(m+1)$, as was the case for $l_A\leq 2$ for
the $m=3$ case.  For $l_A \leq \hf(m+1)$, the basis state is
[0,0,0,...,0], [1,0,0,...,0], [0,1,0,...,0],... [0,0,0,...,1].  The
reduced density matrix is diagonal.  In the thermodynamic limit the
eigenvalues are $(m-l_A)/m$ for the first state (no particles in $A$)
and $1/m$ for each of the rest.  The entanglement entropy is
\[
S_{l_A} ~=~ - \lba\frac{m-l_A}{m}\rba\log\lba\frac{m-l_A}{m}\rba 
~-  l_A \lba \frac{1}{m} \log\frac{1}{m} \rba
\]
which takes the form $S_{l_A}\sim (\log{m}/m)l_A$ for $l_A \ll m$.  
Thus for large $m$ the $S_{l_A}$ versus $\sqrt{l_A}$ curve starts out
quadratic, and only later displays the 
asymptotic linear behavior, with a crossover presumably around $l_A
\sim{m}$.

\begin{figure}
\centering
 \includegraphics*[width=0.95\columnwidth]{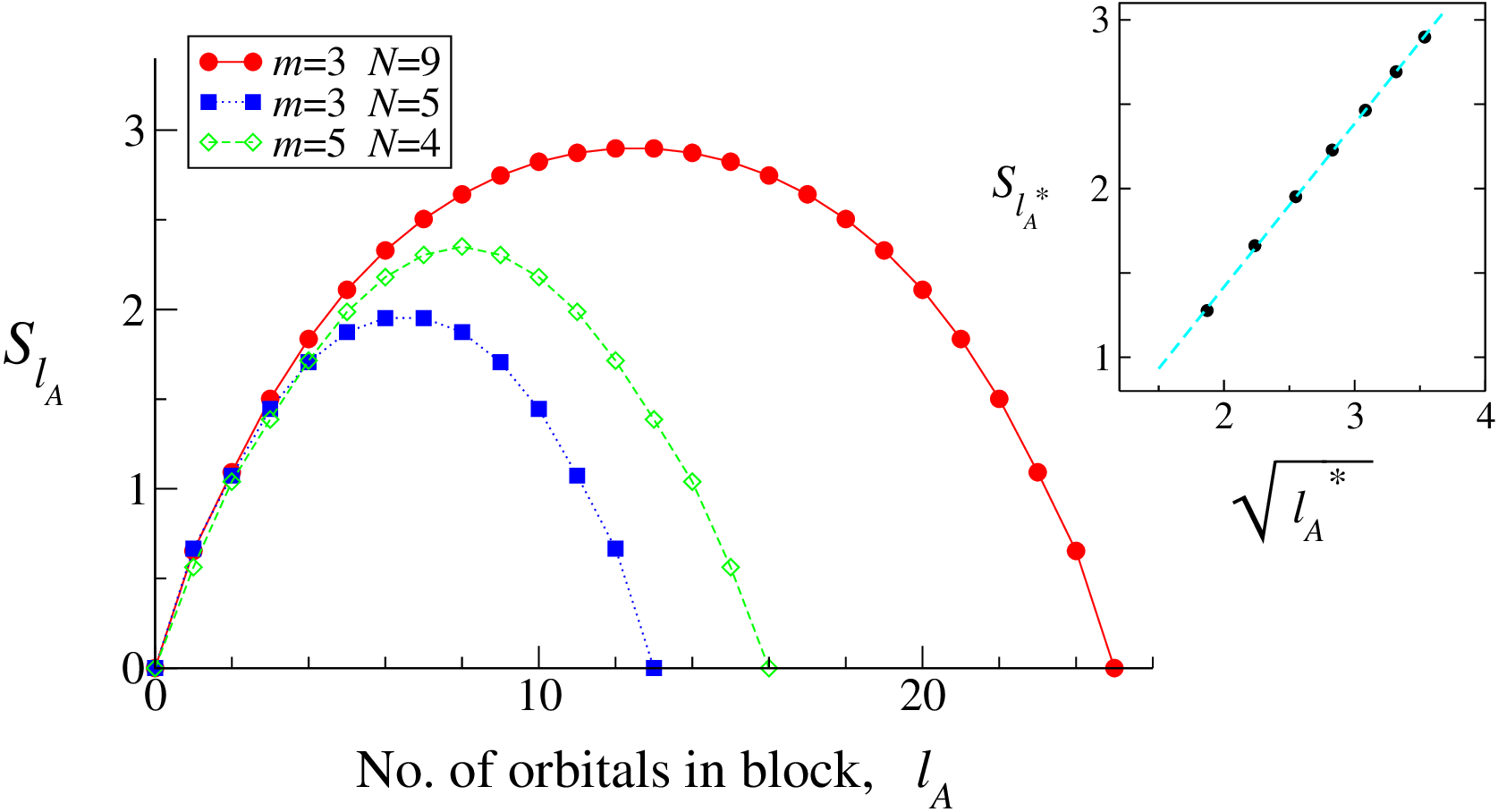} 
\caption{  \label{fig_finiteN-orbital-Ent-variousNm}
(Color online.)  Entanglement entropies with orbital partitioning
  ($S_{l_A}$), for various $N$ and $m$.  Inset: maximum $S_{l_A}$
  values plotted against square root of positions $l_A^*$ of the
  maximum, for $m=3$ wavefunctions of various sizes.
}
\end{figure}

\emph{Orbital partitioning; numerical results} ---
We now present numerical results for entanglement entropy with orbital
partitioning.  The numerical calculation is based on Laughlin
wavefunctions evaluated \emph{exactly} using algebraic methods; some
details are given later in the article.

Fig.~\ref{fig_finiteN-orbital-Ent-variousNm} shows numerically
calculated orbital entanglements ($S_{l_A}$) for several of our
wavefunctions, as a function of the number of orbitals ($l_A$) in the
$A$ partition. 
The partition flip symmetry of the entanglement entropy ($S_A=S_B$)
implies that $S_{l_A}$ is arc-shaped with a maximum at
$l_A^*=(N_{\phi}+1)/2$.
The initial increasing parts of these curves reflect physics of the
macroscopic state, while the downward curvature is a finite-size
effect.  Scaling information is not immediately obvious because the
curves bend down before reaching large $l_A$.  However, we can extract
some information about the thermodynamic limit
by plotting the maximum $S_{l^*_A}$ of the arc-shaped curves against
the position of the maxima, $l_A^*$.
While the detailed finite-size effects are difficult to analyze, it is
reasonable to assume that the macroscopic functional dependence of
$S_{l_A}$ (e.g., algebraic or logarithmic dependence) will also appear
in the dependence of the maximum.
%
%
The inset to Fig.~\ref{fig_finiteN-orbital-Ent-variousNm} shows such a
plot for $m=3$ states, with ${l_A}$'s plotted in a square root scale.
The linear behavior is manifest.  We have observed the same feature
for $m=5$ wavefunctions.  We have thus already verified the boundary
law $S_{l_A} \propto \sqrt{l_A}$.

\begin{figure}
\centering
 \includegraphics*[width=0.95\columnwidth]{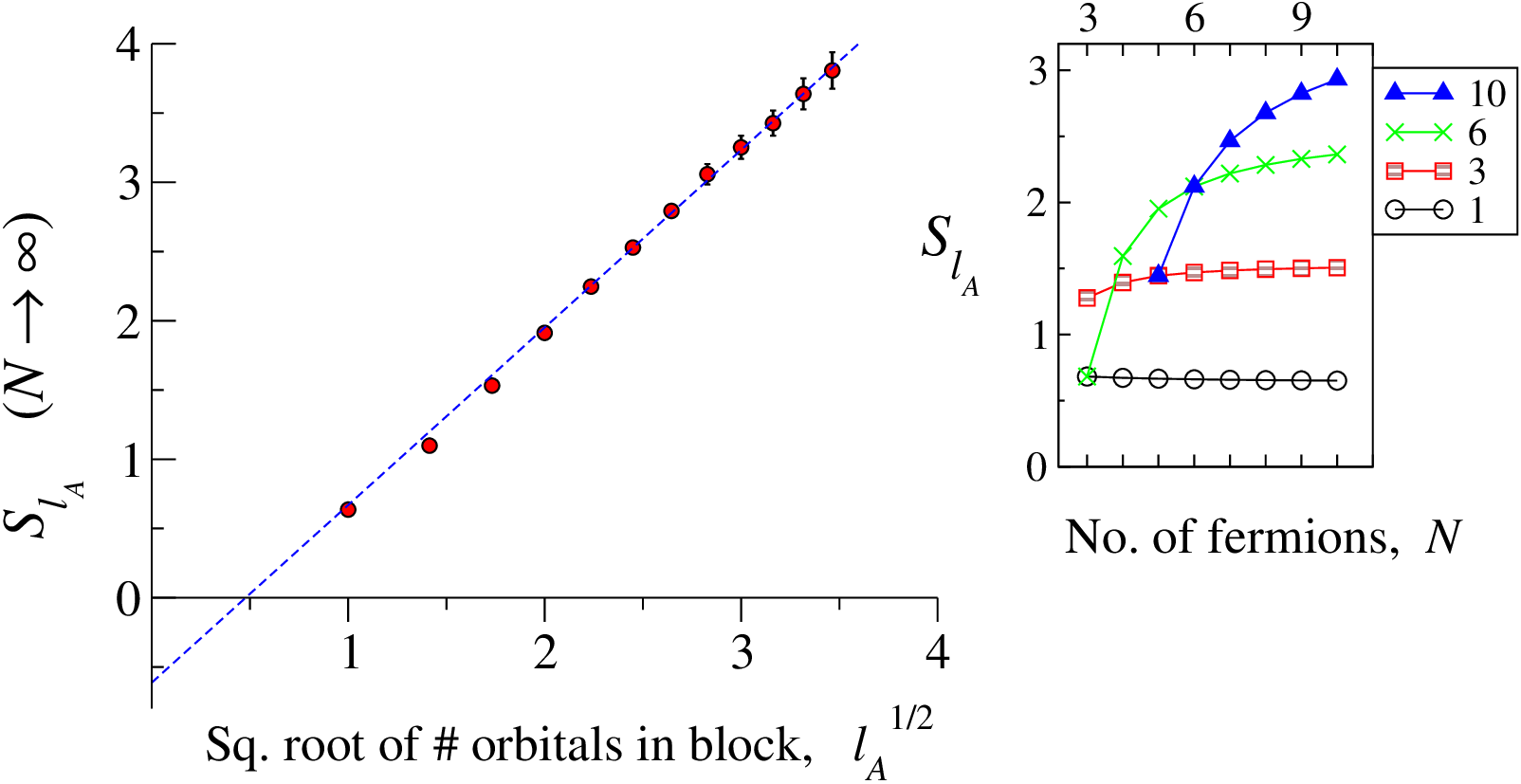} 
\caption{  \label{fig_extrapolatn-m3}
(Color online.)  Entanglement entropies in $m=3$ states extrapolated
  to the thermodynamic limit.  Dashed line is a fit to
  $-\gamma+c_1\sqrt{l_A}$ giving more weight to the higher points.
%
  Inset on right plots $S_{l_A}$ against $N$, for $l_A = 1,3,6,10$.
  With wavefunctions available upto $N=10$, the $N\rightarrow\infty$
  extrapolation is reliable for small $l_A$ but is already difficult
  for $l_A=10$ because the $S_{l_A}(N)$ curve is not nearly flat yet.
}
\end{figure}

A more quantitative approach to the thermodynamic limit is to do a
numerical $N\ra\infty$ extrapolation of the $S_{l_A}$ data for each
$l_A$ (Fig.~\ref{fig_extrapolatn-m3}).  We fit the $S_{l_A}(N^{-1})$
data points (inset to Fig.~\ref{fig_extrapolatn-m3}) to functions like
$c_0 + c_1/N^{\alpha_1} + c_2/N^{\alpha_2}$, noting that
$S_{{l_A}=1,2}$ indeed have expansions for this form.  We use various
sets \{$\alpha_i$\} of small integers, dropping combinations that give
extrapolation functions with pathological features at small $N^{-1}$.
The set of values obtained by this procedure yields the estimates and
errors of Fig.~\ref{fig_extrapolatn-m3}.

\emph{Extracting the topological entropy} ---
We note that the $-\gamma + c_1\sqrt{l_A}$ behavior
\cite{Preskill-Kitaev_PRL06, Levin-Wen_PRL06} is expected only for
$l_A\gtrsim 3$.  For extracting $\gamma$, we thus drop two or more of
the lowest points from the available $S_{{l_A}}$ values.  It is not
feasible to use only the largest values because the extrapolation
uncertainty is largest there (Fig.~\ref{fig_extrapolatn-m3}).
Dropping 2 to 5 of the lowest points, and giving various relative
weights to the higher or lower values, we extract various estimates
for $\gamma$ by fitting $S_{l_A} = -\gamma + c_1\sqrt{l_A}$.  The
resulting estimate for the topological entanglement entropy of the
$m=3$ state is $\gamma = -0.60\pm 0.13$.  This is consistent with the
theoretical value of $-0.55$.  However, larger wavefunctions are
probably required for precisely extracting $\gamma$ for models where
it is not known \emph{a priori}.

\begin{figure}
\centering
 \includegraphics*[width=0.95\columnwidth]{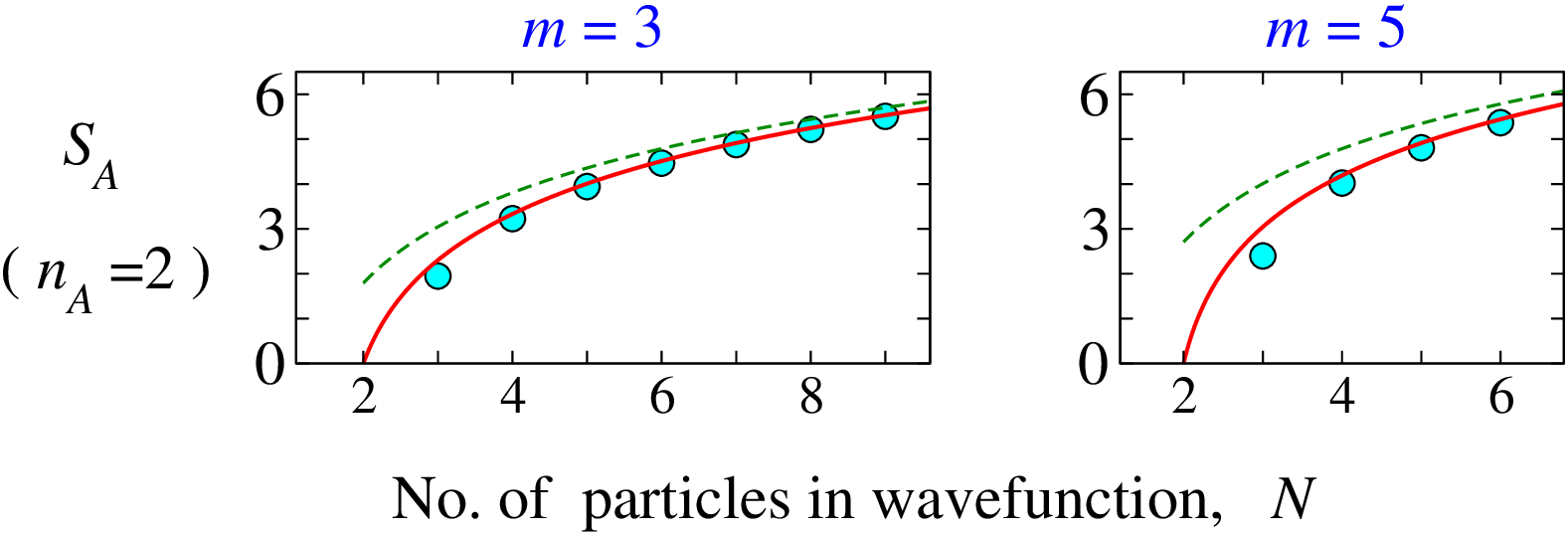} 
\caption{  \label{fig_2-pcle}
(Color online.)  Entanglement entropy of two partices ($A$) with the
  rest ($B$).  Dots are numerical exact values; dashed curves are
  na\"ive estimates [Eq.~\eqref{eq_obvious-bound}]; full curves are
  improved bounds taking multiplet structure into account.
}
\end{figure}

\emph{Particle Entanglement} ---
We now turn to partitioning the Laughlin states into two sets of
particles rather than two blocks of orbitals.  
Noting that the entanglement entropy $S_A$ is largest
and equal to the logarithm of the number of nonzero eigenvalues when
$\rho_A$ has equal nonzero eigenvalues, we can set a na\"ive upper
bound for $S_A$.  When $A$ contains $n_A$ particles out of a total of
$N$ particles in the $\nu=1/m$ Laughlin state, the size of the density
matrix is set in first instance by the number of possible ways in
which the $n_A$ particles can be distributed among $N_{\phi}+1 =
m(N-1)+1$ orbitals; thus
\begin{equation} \label{eq_obvious-bound}
S_A \leq \log \begin{pmatrix}  N_{\phi}+1 \\ n_A \end{pmatrix}  \; .
\end{equation}
This bound becomes an exact result for the integer quantum Hall state
$m=1$, i.e., the noninteracting case where the Laughlin wavefuction
reduces to a single Slater determinant.  For the fractional quantum
Hall states ($m \geq 3$), the bound is sharp for $n_A$=1 but not for
$n_A>1$ (Fig.~\ref{fig_2-pcle}).

We note that the absence of terms in the Laughlin wavefunctions with
adjacent orbital occupancies near the poles implies that, for $n_A>1$,
some of the eigenvalues of the reduced density matrix are zero,
leading to a tighter upper bound than above. This idea becomes much
more effective with the additional observation that the eigenvalues of
$\rho_A$ are organized in a $SU(2)$ multiplet structure, due to
$\rho_A$ commuting with the total angular momentum magnitude and
$z$-component (${\bf L}_{n_A}^2$ and $L_{n_A}^z$) of the $n_A$
particles in $A$.  When any eigenvalue of $\rho_A$ vanishes due to the
absence of corresponding states in the Laughlin wavefunction, the
multiplet structure implies that \emph{every} eigenvalue in the same
multiplet must also vanish.

For $n_A=2$, $m=3$, one observes that the vanishing states are part of
the $L_2 = m(N-1)-1 = N_{\phi}-1$ representation of the $SU(2)$\
symmetry algebra.  This reduces by $2L_2+1 = 2N_{\phi}-1$ the number
of 2-particles states contributing to the entropy.  For $n_A=2$ and
general $m$, there are $(m-1)/2$ multiplets contributing zero
eigenvalues, leading to a total of $\hf(m-1)(2N_{\phi}-m+2)$ zeros.  The
number of nonzero eigenvalues is 
\[
\begin{pmatrix}  N_{\phi}+1 \\ 2 \end{pmatrix}  -\hf(m-1)(2N_{\phi}-m+2)  
=  \begin{pmatrix}  N_{\phi}+1-(m-1) \\ 2 \end{pmatrix}
\]
and the logarithm gives a much better upper bound to the $n_A=2$
entanglement entropies (Fig.~\ref{fig_2-pcle}).  This bound is still
not sharp because the eigenvalues are not all equal.  The eigenvalue
distribution is itself interesting and may be discussed in future
work; however, the effect on $S_A$ is small.  In Fig.~\ref{fig_2-pcle}
we show that the exact values for $S_{n_A=2}$ rapidly converge to our
improved upper bound.

Generalizing to $n_A>2$, we have rigorously established the following
upper bound
\begin{equation}  
\label{eq_pcle-bound}
S_A \leq \log \begin{pmatrix}  N_{\phi}+1-(m-1)(n_A-1) \\ n_A
\end{pmatrix} \ .
\end{equation}
We claim that the entropy $S_A$ will be close to this bound for $n_A
\ll N$.
The derivation of Eq.~(\ref{eq_pcle-bound}) relies on a result of
Ref.~\cite{Read-Rezayi96} for counting degeneracies of Laughlin states
on the sphere in the presence of excess flux, that is, of a number of
quasi-holes. The combinatorial factor in Eq.~(\ref{eq_pcle-bound}) has
the following interpretation: it is the number of ways $n_A$ particles
can be put into $N_{\phi}+1$ orbitals, such that between two occupied
orbitals there are at least $m-1$ that are empty.  Clearly, this
counting represents a version of Haldane's notion of exclusion
statistics \cite{Haldane-exclusion_PRL91}.

\emph{Numerical method} ---
Since it is not known which terms in Laughlin wavefunctions are more
important for various entanglement entropies, we avoided truncation
schemes and calculated \emph{complete} wavefunctions.  Such
calculations are in the spirit of combinatorial considerations of
Refs.~\cite{Dunne_IJMP93, Wavefunctn_algebra-papers}, rather than that
of ``exact diagonalization'' studies.  The enumeration of Laughlin
wavefunction terms and calculation of coefficients are combinatorially
explosive problems \cite{Dunne_IJMP93,Wavefunctn_algebra-papers};
symbolic expansion is not feasible beyond more that 5 or 6 particles.
Instead, we made use of results in Ref.~\cite{Dunne_IJMP93} where the
coefficients are expressed as expansions in characters of the
symmetric group,
$a_{(\mu)}=\sum_{(\mu)}c_{(\lambda)}\chi_{(\mu)}^{(\lambda)}$.  Here
$(\mu)$'s are integer partitions representing terms in the Laughlin
wavefunction.  The conjugacy classes $(\lambda)$ and the coefficients
of expansion $c_{(\lambda)}$ are both found by expanding out a Hankel
determinant symbolically \cite{Dunne_IJMP93}.  Characters of the
symmetric group {\bf S}$_{\hf{m}N(N-1)}$ were calculated using
publicly available group theory code while the Hankel determinant
expansion was done with a symbolic manipulation routine coded from
scratch.  The admissible partitions $(\mu)$ for Laughlin state terms
were generated by combinatorial rules
\cite{Dunne_IJMP93,Wavefunctn_algebra-papers}.
With some effort (50+ cpu-days for the largest wavefunctions), we
calculated $m=3$ wavefunctions up to $N=10$ fermions and $m=5$
wavefunctions up to $N=6$ fermions.  


\emph{Concluding remarks} ---
We conclude that the entanglement entropy pertaining to the two
types of partitioning (spatial and particle) reveal
different aspects of the intricate topological order
of the Laughlin states. This beautifully illustrates that,
more generally, entanglement entropy is a valuable probe for
quantum order. For the case of particle partitioning we remark
the following. Trusting that eq. (2) gives a close bound to
$S_A$, we have for $n_A \ll N$
\[
S_A - \log \begin{pmatrix}
N_\phi+1\\n_A
\end{pmatrix}
\sim  ~-~ \frac{1}{N} \frac{m-1}{m}n_A(n_A-1)
~+~  \ord(1/N^2)
\]
We propose that, in general, the difference between $S_A$ and the
na\"ive expression \eqref{eq_obvious-bound} based on the number of
1-particle states allows a $1/N$ expansion that reveals correlations
and quantum order in a state.

Our work opens up a number of questions.  It would be interesting to
obtain numeric data for $n_A\geq3$ particle entanglement (not
available at present) to investigate how well our bound
\eqref{eq_pcle-bound} works.  With further effort, some larger
wavefunctions may also be obtained, perhaps leading to more precise
extraction procedures for the total quantum dimension. 
%
%
Obviously, other quantum Hall states, e.g., those with
paired or clustered structure and non-abelian statistics, deserve
attention, and further work in this direction is planned.

While writing up our results, we learned of parallel work
\cite{Latorre_FQHE} that has overlap with some issues we consider.

\emph{Acknowledgements} ---
We are grateful for helpful discussions with P.~Calabrese,
R.~Hagemans, V.~Korepin, and K.~Shtengel.  
Funding was provided by the Stichting voor Fundamenteel Onderzoek der
Materie (FOM) and the Nederlandse Organisatie voor Wetenschaplijk
Onderzoek (NWO).


\end{document}